\documentclass[draft]{agujournal2019}
\usepackage{url} 
\usepackage{lineno}
\usepackage[inline]{trackchanges} 
\usepackage{soul}

\usepackage{amsmath}

\draftfalse

\journalname{arXiv}

\begin{document}

\title{Linking distributed and integrated fiber-optic sensing}

\authors{Daniel C. Bowden\affil{1}, Andreas Fichtner\affil{1}, Thomas Nikas\affil{2}, Adonis Bogris\affil{3}, Christos Simos\affil{4}, Krystyna Smolinski\affil{1}, Maria Koroni\affil{1}, Konstantinos Lentas\affil{5} , Iraklis Simos\affil{6},  Nikolaos S. Melis\affil{5} }
\affiliation{1}{Institute of Geophysics, ETH Z{\"u}rich, 8092, Z{\"u}rich, Switzerland}
\affiliation{2}{Dept. of Informatics and Telecommunications, National and Kapodistrian University of Athens, Athens, GR-15784, Greece }
\affiliation{3}{Dept. of Informatics and Computer Engineering, University of West Attica, Aghiou Spiridonos, 12243, Egaleo, Greece}
\affiliation{4}{University of Thessaly, Dept. of Physics, Electronics and Photonics Laboratory, 35100 Lamia, Greece }
\affiliation{5}{National Observatory of Athens, Institute of Geodynamics, Greece }
\affiliation{6}{Department of Electrical and Electronics Engineering, University of West Attica, Aghiou Spiridonos, 12243, Egaleo, Greece}
\correspondingauthor{Daniel C. Bowden}{daniel.bowden@erdw.ethz.ch}

\begin{keypoints}
\item We develop and test a theory for the quantitative comparison of distributed and integrated fiber-optic strain sensing

\item The sensitivity of integrated measurements depends primarily on local fiber curvature and fiber heterogeneity

\item A data-based comparison corroborates the theory, thereby demonstrating the geophysical value of our integrated sensor system
\end{keypoints}

%
%

\begin{abstract}
Distributed Acoustic Sensing (DAS) has become a popular method of observing seismic wavefields: backscattered pulses of light reveal strains or strain-rates at any location along a fiber-optic cable. In contrast, a few newer systems transmit light through a cable and collect integrated phase delays over the entire cable, such as the Microwave Frequency Fiber Interferometer (MFFI). These integrated systems can be deployed over significantly longer distances, may be used in conjunction with live telecommunications, and can be significantly cheaper. However, they provide only a single time series representing strain over the entire length of fiber. This work discusses theoretically how a distributed and integrated system can be quantitatively compared, and we note that the sensitivity depends strongly on points of curvature. Importantly, this work presents the first results of a quantitative, head-to-head comparison of a DAS and the integrated MFFI system using pre-existing telecommunications fibers in Athens, Greece. \end{abstract}

\section*{Plain Language Summary}
New technologies are being developed to measure earthquakes using fiber-optic telecommunications cables. The most popular new method in recent years is "Distributed Acoustic Sensing," (DAS) in which pulses of light are repeatedly sent down a fiber and one measures the signals that reflect back. Shaking from an earthquake will stretch the fiber and the signature of reflected pulses will change. A new method (MFFI in the paper) sends light from one end to the other and measures differences in optical phase. The new method has many advantages: it is cheaper, can be used on longer cables, and can be used at the same time as active telecommunications. As a disadvantage is lacks the high spatial resolution that DAS offers. This paper discusses differences between the two methods and shows how to compare them, and then shows the first real-data, head-to-head comparison from an earthquake observed in Athens, Greece.

%
%
\section{Introduction}
Distributed Acoustic Sensing (DAS) has grown in application in recent years, as a method of measuring strains of a seismic wavefield. By sending pulses of light through a fiber-optic cable and measuring the changing signature of backscattered light, a single DAS interrogator can effectively measure strains or strain rates at thousands of locations along the fiber. Dense channel spacing on a meter scale and high sampling frequencies have made it attractive for seismological studies, including earthquake and aftershock monitoring \cite{Nayak2021,Li2021}, fault-zone imaging \cite{Jousset2018,Lindsey2019}, in boreholes \cite{Lellouch2019}, on glaciers \cite{Walter2020}, on volcanoes \cite{Klaasen2021,Currenti2021}, and many others. Many such studies have exploited pre-existing telecommunications infrastructure, or ``dark fibers'' not in use; this potential has further enabled DAS observations in dense urban areas where characterizing seismic hazard is particularly important \cite{Biondi2017,Martin2018,Yuan2020}. It is precisely these urban areas where deploying new seismic instrumentation is challenging, and yet dense site-effect studies and microseismic monitoring is crucial. \\
Typical DAS systems are limited in the distance of fiber usable, however, listed by manufacturers usually in the 10s of kilometers. There are limitations on how far the light can propagate before the backscattered signal is too attenuated and weak. Also a limiting factor is how long it takes for light to propagate back -- a longer two-way travel time will mean the system has to wait longer before sending the next pulse, limiting the maximum sampling frequency. \\
In contrast, various systems have been proposed based on transmission of light through a fiber, measuring the signal at the end (or after being looped back to the start). Such systems might exploit considerably longer fibers, offering seismologists a way to use existing transoceanic cables. Some authors have successfully measured polarization changes accumulated along the entire fibers \cite{Mecozzi2021,Zhan2021}, while others might measure changes in phase \cite{Marra2018}. Recently, \citeA{Bogris2021,Bogris2022} installed one such system in a suburban region of Athens, using pre-existing telecommunications fiber. Their particular system relies on the interferometric use of microwave-range frequencies -- signals sent along the fiber and back in a closed loop are compared to what was sent and phase differences are measured. As with DAS, repeated observations allow them to measure changing strains along the fiber in real-time. Referred to as a Microwave Frequency Fiber Interferometer (MFFI), the system is a fraction of the cost of a typical DAS interrogator and can be used in parallel with live telecommunications signals. \\
A notable difference between the MFFI system (and all such direct transmission systems) as compared to DAS: the resulting observation is a single measure of strain integrated along the entire fiber length. This makes the measurement potentially harder for seismologists to interpret. This paper theoretically outlines how a local strain wavefield is observed on such a system, and discusses how local earthquakes may or may not be seen in different circumstances. This sensitivity depends, namely, on fiber curvature; straight segments of fiber effectively average out to give very little contribution. \\
In September and October of 2021, we ran a Silixa iDAS interrogator alongside the MFFI system of \citeA{Bogris2021,Bogris2022} in northern Athens, Greece, in collaboration with the Hellenic Telecommunications Organisation (in Greek: OTE). A number of earthquakes were observed, including both local- and regional-scale events. We propose a way to integrate the many thousands of densely spaced DAS strain-rate channels and thus recover what the MFFI system would observe, and subsequently find that the two systems agree remarkably well. This first quantitative comparison of integrated and distributed fiber-optic sensing validates the systems' outputs, and solidifies our understanding of how physical wavefields in the Earth will be measured by emerging integrated sensing systems.\\
%

\section{Theoretical background of phase transmission fibre optics}

This section focuses on the interpretation of a strain wavefield, as measured by a system based on transmitted phase changes, including those of \citeA{Marra2018} and \citeA{Bogris2021,Bogris2022}. We do not focus on the technical or engineering details, and instead analyse how a seismic wavefield is represented and interpreted. More details and toy examples can be found in \citeA{Fichtner2022}.

\subsection{Exact relations between fiber deformation and optical phase changes}

\begin{center}
\begin{figure}
\center\scalebox{0.48}{\includegraphics{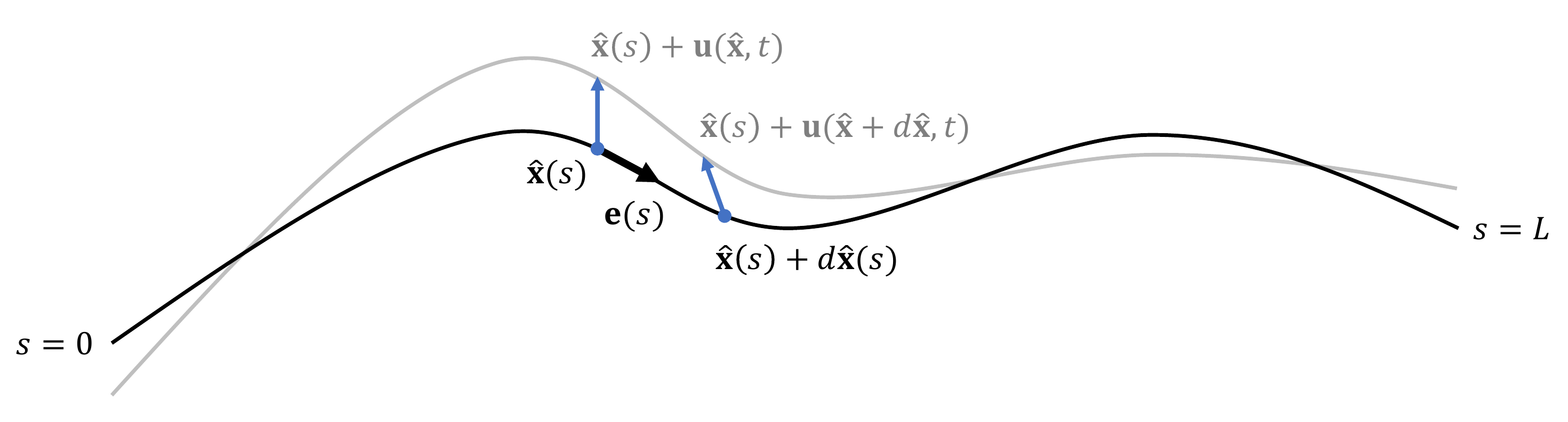}}
\caption{Schematic illustration of fiber deformation. The undeformed fiber, shown as black curve, is represented by the position vector $\hat{\mathbf{x}}(s)$, which is parametrized in terms of the arc length $s$. The cable starts at $s=0$ and ends at $s=L$. Under a displacement field $\mathbf{u}(\hat{\mathbf{x}},t)$, displayed as blue arrows, the Lagrangian position $\hat{\mathbf{x}}(s)$ along the undeformed fiber moves to $\hat{\mathbf{x}}(s)+d\hat{\mathbf{x}}(s)$. The result is the deformed fiber, shown in grey. The local tangent vector $\mathbf{e}(s)$ is shown as a thick black arrow.}\label{F:geometry}
\end{figure}
\end{center}

We begin with the derivation of an exact relation between the deformation tensor $\mathbf{F}(\mathbf{x},t)$ along the fiber and the time-dependent phase $\phi(t)$ of a pulse that propagates from the beginning to the end of the fiber. The only assumption is that the traveltime of the pulse is much smaller than the characteristic time scales of deformation, meaning that the fiber does not deform significantly while a pulse is propagating. To ease calculations, we adopt a parameterized representation of the fiber, with its position $\hat{\mathbf{x}}(s)$ given in terms of the arc length $s$. The latter ranges between $0$ and the total length of the fiber $L$, as shown in Fig. \ref{F:geometry}. 
%
%
In the undeformed state, the time it takes for a pulse to travel from fiber location $\hat{\mathbf{x}}(s)$ to the neighbouring location $\hat{\mathbf{x}}(s)+d\hat{\mathbf{x}}(s)$ is given by $dT = |d\hat{\mathbf{x}}(s)|/c[\hat{\mathbf{x}}(s)]$, where $c[\hat{\mathbf{x}}(s)]$ is the potentially space-dependent speed of light along the fiber. By definition of the arc length, we can express the total traveltime of the pulse as
\begin{linenomath*}
\begin{equation}\label{E:002}
T = \int_{s=0}^L \frac{ds}{c[\hat{\mathbf{x}}(s)]}\,.
\end{equation}
\end{linenomath*}
Under deformation, position $\hat{\mathbf{x}}$ moves to  $\hat{\mathbf{x}}+\mathbf{u}(\hat{\mathbf{x}},t)$, where $\mathbf{u}(\hat{\mathbf{x}},t)$ is the (seismic) displacement field, as illustrated in Fig. \ref{F:geometry}. The neighboring point at the original position $\hat{\mathbf{x}}+d\hat{\mathbf{x}}$ moves to $\hat{\mathbf{x}}+d\hat{\mathbf{x}}+\mathbf{u}(\hat{\mathbf{x}}+d\hat{\mathbf{x}},t)$. It follows that the traveltime of the pulse within the deformed segment of the cable is now
\begin{linenomath*}
\begin{equation}\label{E:100}
dT(t) = \frac{|d\hat{\mathbf{x}}+\mathbf{u}(\hat{\mathbf{x}}+d\hat{\mathbf{x}},t)-\mathbf{u}(\hat{\mathbf{x}},t)|}{c[\hat{\mathbf{x}},\mathbf{u}(\hat{\mathbf{x}},t)]}\,.
\end{equation}
\end{linenomath*}
The denominator accounts for the photo-elastic effect, i.e., changes in the speed of light induced by deformation. Since $d\hat{\mathbf{x}}$ is infinitesimally small, we can rewrite the numerator as
\begin{linenomath*}
\begin{equation}\label{E:101}
\mathbf{u}(\hat{\mathbf{x}}+d\hat{\mathbf{x}},t)-\mathbf{u}(\hat{\mathbf{x}},t) = \mathbf{F}(\hat{\mathbf{x}},t)\, d\hat{\mathbf{x}}\,,
\end{equation}
\end{linenomath*}
where the components of the deformation tensor $\mathbf{F}$ are defined by $F_{ij} = \partial u_i/\partial x_j$. In terms of $\mathbf{F}$, we can rewrite (\ref{E:100}) as
\begin{linenomath*}
\begin{equation}\label{E:103}
dT(t) = \frac{|d\hat{\mathbf{x}}+\mathbf{F}(\hat{\mathbf{x}},t)\, d\hat{\mathbf{x}}|}{c[\hat{\mathbf{x}},\mathbf{u}(\hat{\mathbf{x}},t)]}\,.
\end{equation}
\end{linenomath*}
This can be further simplified using the arc-length parametrization of the position vector, $d\hat{\mathbf{x}} = \mathbf{e}(s)\, ds$, where $\mathbf{e}(s)$ is the normalized tangent vector along the fiber. With this, we find
\begin{linenomath*}
\begin{equation}\label{E:105}
dT(t) = \frac{|[\mathbf{I}+\mathbf{F}(\hat{\mathbf{x}},t)]\,\mathbf{e}(s)|}{c[\hat{\mathbf{x}},\mathbf{u}(\hat{\mathbf{x}},t)]} ds\,,
\end{equation}
\end{linenomath*}
and the total, time-dependent traveltime of the pulse becomes
\begin{linenomath*}
\begin{equation}\label{E:106}
T(t) = \int_{s=0}^L \frac{|[\mathbf{I}+\mathbf{F}(\hat{\mathbf{x}},t)]\,\mathbf{e}(s)|}{c[\hat{\mathbf{x}},\mathbf{u}(\hat{\mathbf{x}},t)]} ds\,.
\end{equation}
\end{linenomath*}
In the specific case of a monochromatic input with circular frequency $\omega$, the traveltime difference $\Delta T(t) = T(t) - T$ translates into a phase difference $\phi(t) = \omega \Delta T(t)$ between the reference and the deformed state. Substituting (\ref{E:002}) and (\ref{E:106}), and taking the time derivative, we obtain the phase changes with respect to time,
\begin{linenomath*}
\begin{equation}\label{E:202}
\partial_t\phi(t) = \omega \partial_t \int_{s=0}^L \frac{|[\mathbf{I}+\mathbf{F}(\hat{\mathbf{x}},t)]\,\mathbf{e}(s)|}{c[\hat{\mathbf{x}},\mathbf{u}(\hat{\mathbf{x}},t)]} ds\,.
\end{equation}
\end{linenomath*}
Eq. (\ref{E:202}) is valid without any approximations, and it relates measured phase changes of the monochromatic laser signal to the deformation field $\mathbf{u}(\hat{\mathbf{x}},t)$ along the fiber.

\subsection{First-order approximations and relations to Distributed Acoustic Sensing (DAS)}

Eq. (\ref{E:202}) can be simplified considerably by realizing that typical seismic displacement fields $\mathbf{u}$ have amplitudes in the nano- or micrometer range. Therefore, the norm of the deformation tensor $\mathbf{F}$ is typically orders of magnitude smaller than $1$. It follows that first-order approximations can easily be justified. To avoid clumsy notation, we work with a slight reformulation of Eq. (\ref{E:202}), which uses the refractive index $n=c_0/c$, where $c_0$ is the speed of light in vacuum. To simplify (\ref{E:202}), we employ the first-order approximation
\begin{linenomath*}
\begin{equation}\label{E:300}
|[\mathbf{I}+\mathbf{F}(\hat{\mathbf{x}},t)]\,\mathbf{e}(s)|^2 \approx 1 + 2\mathbf{e}^T \mathbf{E} \mathbf{e}\,,
\end{equation}
\end{linenomath*}
with the strain tensor $\mathbf{E}=(\mathbf{F}^T + \mathbf{F})/2$. Denoting the strain along the fiber as $\epsilon=\mathbf{e}^T \mathbf{E} \mathbf{e}$ and using the first-order relation $\sqrt{1+2\epsilon} \overset{.}{=} 1 + \epsilon$, we arrive at
\begin{linenomath*}
\begin{equation}\label{E:301}
\partial_t\phi(t) \approx \frac{\omega}{c_0}\, \partial_t \int_{s=0}^L n[\hat{\mathbf{x}},\mathbf{u}(\hat{\mathbf{x}},t)] \, \left( 1 + \epsilon[\hat{\mathbf{x}}(s),t] \right)\, ds\,.
\end{equation}
\end{linenomath*}
Assuming that $n$ primarily depends on $\epsilon$, we may use the first-order Taylor expansion $n(\epsilon)\overset{.}{=}n_0 + n'\epsilon$, thereby obtaining
\begin{linenomath*}
\begin{equation}\label{E:302}
 \partial_t\phi(t) \approx \frac{\omega}{c_0}\, \partial_t \int_{s=0}^L \tilde{n}[\hat{\mathbf{x}}(s)] \, \epsilon[\hat{\mathbf{x}}(s),t]\, ds\,, 
\end{equation}
\end{linenomath*}
where $\tilde{n}=n_0 + n'$ is the sum of the static refractive index and the axial strain derivative of the refractive index. The latter is referred to as the photo elastic effect of the material, and has been experimentally shown to contribute only about 1/5 the amplitude compared to length changes of the fiber \cite{Bertholds1988}. Regardless of how that refractive index is measured, Eq. (\ref{E:302}) provides a direct relation between phase changes $\partial_t\phi$ measured by the transmission system, and the axial strain rate $\partial_t\epsilon$. In the case where the refractive index is roughly constant along the fiber, it suffices to integrate DAS measurements of $\partial_t\epsilon$ along the fiber in order to synthesize transmission measurements of $\partial_t\phi$. In practice, integrating is achieved by simply summing all the individual DAS channels.

\subsection{The role of cable curvature}

To gain deeper understanding of these results, we derive an alternative representation of (\ref{E:302}). For this, we return to Eq. (\ref{E:100}). Denoting by $\hat{\mathbf{u}}(s,t) = \mathbf{u}(\hat{\mathbf{x}}(s),t)$ the displacement field along the fiber, we obtain
\begin{linenomath*}
\begin{equation}\label{E:500}
|d\hat{\mathbf{x}}+\mathbf{u}(\hat{\mathbf{x}}+d\hat{\mathbf{x}},t)-\mathbf{u}(\hat{\mathbf{x}},t)| = |d\hat{\mathbf{x}}+\hat{\mathbf{u}}(s+ds,t)-\hat{\mathbf{u}}(s,t)| = \left| \mathbf{e}(s) + \partial_s \hat{\mathbf{u}}(s,t) \right| ds\,,
\end{equation} 
\end{linenomath*}
where $\partial_s$ denotes the derivative with respect to the arc length $s$. Again making use of the first-order relation
\begin{linenomath*}
\begin{equation}\label{E:501}
\left| \mathbf{e}(s) + \partial_s \hat{\mathbf{u}}(s,t) \right| = 1 + \mathbf{e}(s)^T \partial_s \hat{\mathbf{u}}(s,t)\,,
\end{equation}
\end{linenomath*}
and the Taylor expansion of the refractive index, yields a new version of Eq. (\ref{E:302})
\begin{linenomath*}
\begin{equation}\label{E:502}
\partial_t\phi(t) \overset{.}{=} \frac{\omega}{c_0}\, \partial_t \int_{s=0}^L \tilde{n}[\hat{\mathbf{x}}(s)]\,\mathbf{e}(s)^T \partial_s\hat{\mathbf{u}}(s,t)\, ds\,. 
\end{equation}
\end{linenomath*}
Using the product rule,
\begin{linenomath*}
\begin{equation}\label{E:503}
\tilde{n}[\hat{\mathbf{x}}(s)]\,\mathbf{e}(s)^T \partial_s\hat{\mathbf{u}}(s,t) = \partial_s \left( \tilde{n}[\hat{\mathbf{x}}(s)]\, \mathbf{e}(s)^T  \hat{\mathbf{u}}(s,t) \right) - \partial_s \left( \tilde{n}[\hat{\mathbf{x}}(s)] \, \mathbf{e}(s)^T \right) \, \hat{\mathbf{u}}(s,t)\,,
\end{equation}
\end{linenomath*}
and substituting back into Eq. (\ref{E:302}), we find
\begin{linenomath*}
\begin{equation}\label{E:504}
\partial_t\phi(t) \overset{.}{=} \underbrace{ \frac{\omega}{c_0}\, \partial_t \left. \left( \tilde{n}[\hat{\mathbf{x}}(s)] \, \mathbf{e}(s)^T \hat{\mathbf{u}}(s,t) \right) \right|_{s=0}^{s=L} }_{\text{start/end point contribution}}- \underbrace{ \frac{\omega}{c_0}\, \partial_t \int_{s=0}^L \partial_s \left( \tilde{n}[\hat{\mathbf{x}}(s)]\,\mathbf{e}(s)^T  \right) \, \hat{\mathbf{u}}(s,t)\, ds}_{\text{curvature contribution}}\,.
\end{equation}
\end{linenomath*}
Eq. (\ref{E:504}) has two contributions to the observed phase changes. The first one originates from the displacement at the start and end points of the fiber. It vanishes when the start and end points are not affected by deformation. This may happen in cases of rather localized deformation that only happens along a smaller section of the fiber.\\[5pt]
The second contribution results from changes of the tangent vector and the effective refractive index along the fiber. It follows that strongly curved segments make a larger contribution to the integrated phase change than segments that are nearly straight. Spatial variations in the effective refractive index may result, for example, from non-symmetric, non-circular fiber cores that change along the fiber and induce a birefringence and polarization changes. This effect is frequency-dependent, and results in a Polarization Mode Dispersion (PMD) that is the key observable in the system used by \citeA{Mecozzi2021} and \citeA{Zhan2021}. However, for phase observations of the MFFI or similar systems, this polarization effect is negligible compared to the photo-elastic effect and changing fiber lengths \cite{Butter1978}.\\[5pt]
Finally, the extent to which second-order effects in the Taylor expansion in Eq. \ref{E:504} play a role are explored numerically by \citeA{Fichtner2022} and are shown to be negligible.

%

\section{Experimental comparison of integrated and distributed sensing}

\begin{center}
\begin{figure}
\center{\includegraphics[width=0.95\textwidth]{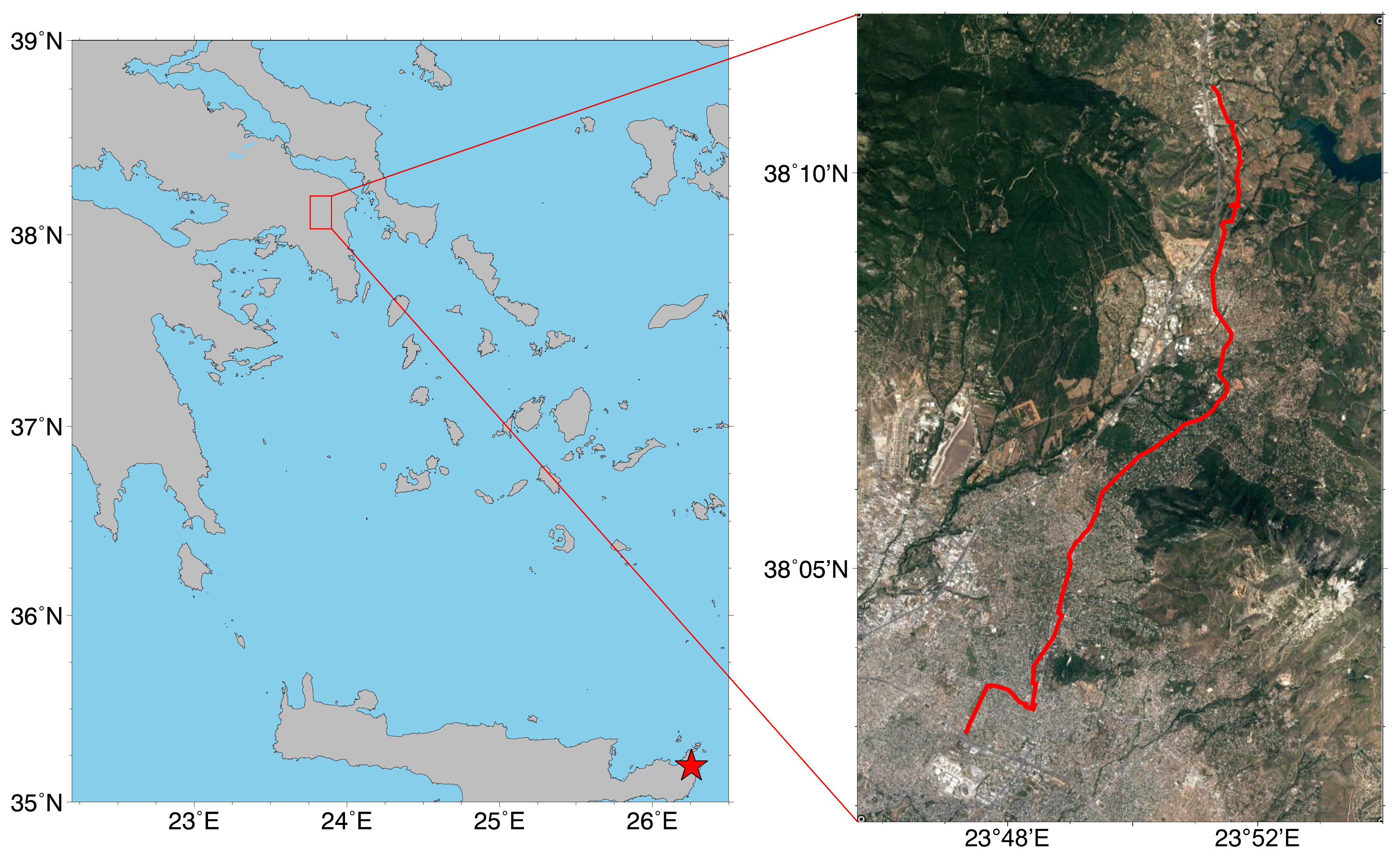}}
\caption{The location of the fiber-optic cables used in Greece. Panel A shows a broad part of Greece and the Aegean Sea, with a red star indicating the epicenter of the M 6.3 earthquake discussed later and a red box indicating the region expanded in the next panel. Panel B shows a region of northern Athens and the cable containing both the DAS and MFFI fibers. \label{F:map}}
\end{figure}
\end{center}

\begin{center}
\begin{figure}
\center{\includegraphics[width=0.95\textwidth]{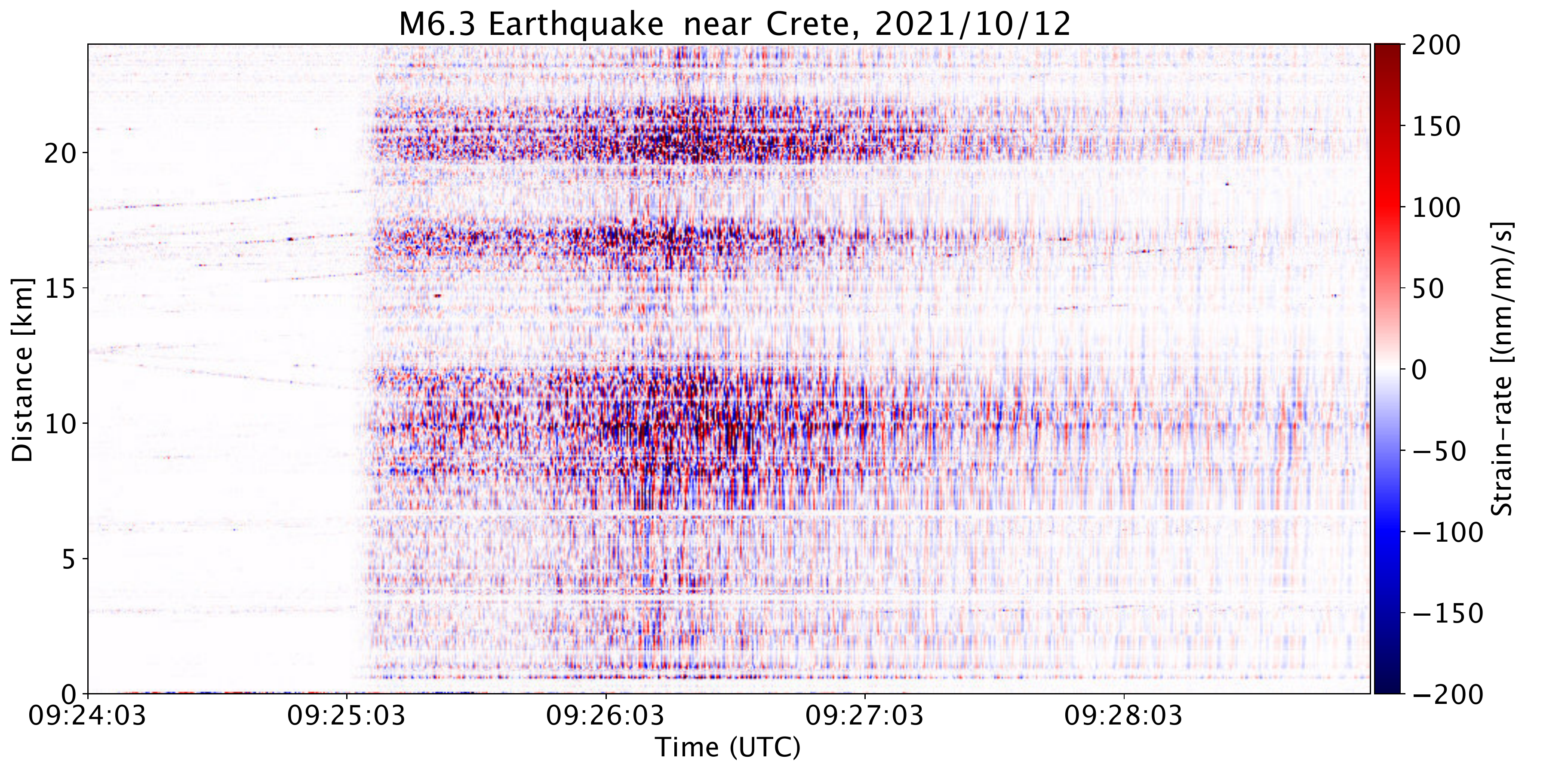}}
\caption{The ML 6.3 Crete earthquake as recorded by the DAS system, recorded with a 2 m channel spacing. The data is broadly filtered between 0.05 Hz and 5 Hz}\label{F:eq1}
\end{figure}
\end{center}

\begin{center}
\begin{figure}
\center{\includegraphics[width=0.95\textwidth]{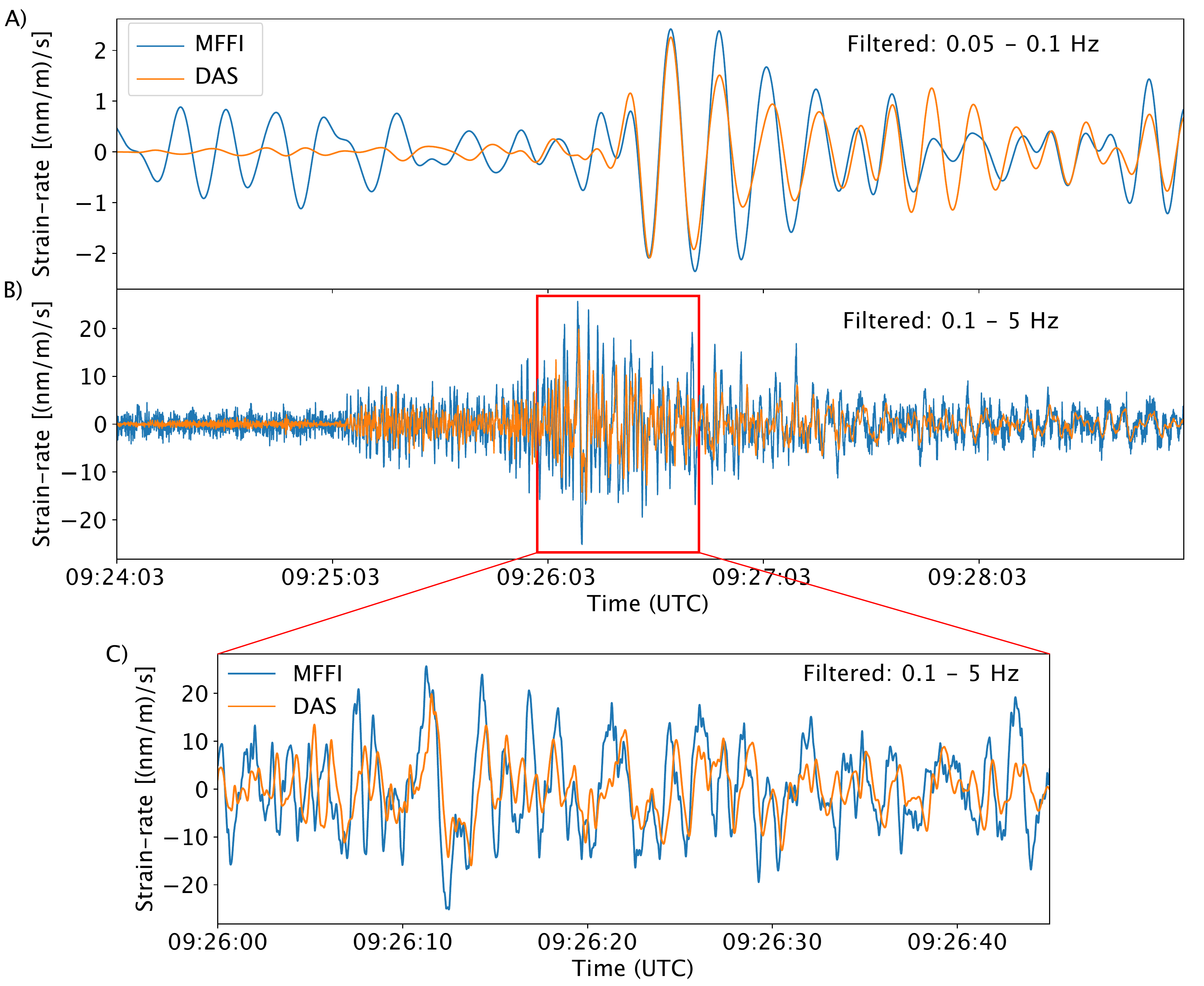}}
\caption{The ML 6.3 Crete earthquake as recorded by both systems. DAS channels are integrated to produce the orange time series, and the MFFI outputs are in blue. Panel A shows signals at a lower frequency range of 0.05 to 0.1 Hz, while Panels B and C show a higher frequency range of 0.1 to 5 Hz. Panel C zooms in on a narrower time range of the signals in Panel B.}\label{F:eq1_compare}
\end{figure}
\end{center}

\begin{center}
\begin{figure}
\center{\includegraphics[width=0.5\textwidth]{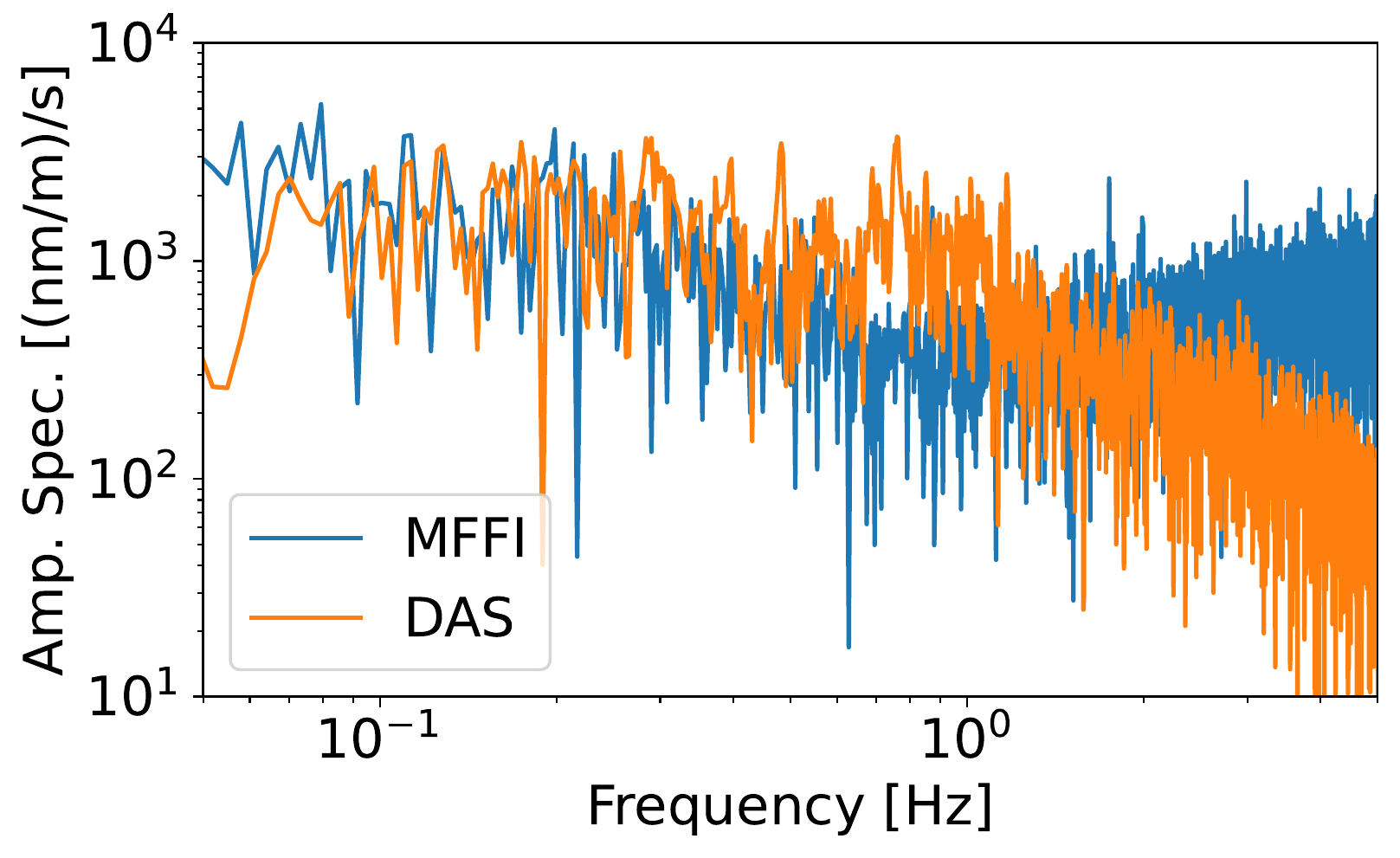}}
\caption{The ML 6.3 Crete earthquake as recorded by both systems, here in frequency domain.}\label{F:eq1_spectra}
\end{figure}
\end{center}

From the theory above (specifically Eq. \ref{E:302}), we see that we can directly compare a DAS system to the MFFI output. In September and October of 2021, we operated a Silixa iDAS interrogator alongside the system described by \citeA{Bogris2021,Bogris2022}. Both systems used pre-existing telecommunications fiber in collaboration with the Hellenic Telecommunications Organisation (OTE), and the work was completed with assistance from the National Observatory of Athens. Fig. \ref{F:map} shows the extent of the fiber, with both the DAS interrogator and the transmission interferometer housed at the OTE Academy building at the southwest end. The total length of fiber used for the DAS measurements was roughly 24 km, while the transmission based system followed the same path and then was looped back to return to the starting location. This extra travel path is accounted for when converting optical phase delays to strain-rate \cite{Bogris2021,Bogris2022}.
%
%
Several earthquakes occurred during the time in which both systems were running. Notably, an earthquake of ML 6.3 occurred on October 12th near the island of Crete, roughly 380 km to the southeast (red star in Fig. \ref{F:map}). Despite the distance, the earthquake resulted in ground motions in Athens as strong as 0.1 cm/s as reported on a nearby strong motion sensor, HL.PLT, operated by the National Observatory of Greece \cite{NOAHL}. \\
Fig. \ref{F:eq1} shows the individual DAS channels along the fiber, while Fig. \ref{F:eq1_compare} shows the comparison between systems. Individual DAS channels in Fig. \ref{F:eq1} are integrated (summed) to produce the orange time series in Fig. \ref{F:eq1_compare}, while the blue time series represents the output from the MFFI system of \citeA{Bogris2022}. We see that the strain rates measured by the two systems agree to within the pre-event noise at both lower frequencies (Panel A) and higher frequencies (Panel B). This includes strong agreement of the timing of seismic phases between the two systems.\\
%
%
%
%
%
The overall amplitude of the averaged DAS signals and the MFFI output are rather smaller than any individual DAS channel. We expect that averaging the many local strains will lead to constructive and destructive interference, since the various phases arrive at different sections of the cable at different times, resulting in a total signal that is lower in amplitude. Fortunately, non-coherent signals such as those resulting from local urban noise and cars are also expected to be relatively suppressed compared to an individaul DAS trace.\\
Also, as with many DAS studies, in Fig. \ref{F:eq1} we observe regions where DAS strain-rates are stronger or weaker. This may be a result of any number of factors, for example from local soil conditions and site amplifications, local coupling of the fiber to the ground, or varying directions of the fiber relative to the incident phases. This does not pose an issue for our comparison of the two systems since the MFFI system uses the same cable, but we note that these sensitivities will be important to characterize if earthquake magnitudes are to be related to strains in the future\cite{Lior2021}.\\
When considering the spectra of both systems, we again see remarkable similarity in reported strain-rates as in Fig. \ref{F:eq1_spectra}. Above 5 Hz, it is expected that the DAS system may begin to be affected by aliasing from the 10 m gauge length \cite{Paitz2021}, or other imperfections from averaging with a 2 meter channel spacing. Additionally, the MFFI system shows a linear increase in system noise at higher frequencies, but this is expected given the limitations of the devices used in that prototype \cite{Bogris2021, Bogris2022}, notably the inexpensive Arduino Analog-to-Digital Converter. Such limitations can be overcome in future iterations.\\
%
%

\section{Discussion and Conclusions}

We developed exact and first-order theories for integrated fiber-optic deformation sensing. The resulting equations enable a quantitative comparison with distributed strain sensing (DAS), and they highlight the dependence of the integrated measurements on fiber curvature and heterogeneity.\\
Parallel experiments for integrated and distributed sensing performed in Athens, using pre-existing telecommunication fibers, allow us to verify the theory with real data and to compare the two systems. The systems report quantitatively similar measurements of strain-rates from a moderately sized, regional earthquake, in both time and frequency domain. Although earlier studies have demonstrated the feasibility of integrated fiber sensing systems \cite{Marra2018,Zhan2021,Mecozzi2021}, this comparison to DAS provides a crucial step for seismologists to understand and quantitatively interpret such observations in the future. \\
However, the fact that the integrated system's sensitivity is highly dependent on curvature means one could miss small ground-motion events. In the case of small events along a straight fiber section, we expect that little to no total strain measurement would appear. Only in cases where such small waves reached a place where the cable bends, or else one of the terminus points, would an integrated strain signal be present. This means care needs to be taken when choosing pre-existing fibers or when designing new systems, or alternatively one needs to at least be aware of the potential spatial gaps in one's coverage. \\
We envision that future seismological studies will exploit a range of complementary sensor types at different scales: DAS or dense geophone arrays for very spatially dense wavefield observations \cite{Muir2021}; existing broadband seismic networks or point-measurement interferometers at the terminal end of a fiber-optic cable \cite{Seat2016} for high-precision observations; and integrated strain sensing systems like MFFI for observations over particularly long distances, hard-to-reach locations (e.g., offshore), or when longer term monitoring is required. It has already been shown that mixed-instrument arrays can be used for even relatively complex, full-waveform analysis \cite{Paitz2019}, and also that integrated strain sensing systems can fit into this same framework for tomography \cite{Fichtner2022b}.\\ 
Integrated strain sensing systems, such as the MFFI system designed by \citeA{Bogris2021,Bogris2022}, are especially exciting in the context of pre-existing telecommunications fibers already connecting our cities. The fact that it can be deployed cheaply and using live telecommunications networks means multiple systems could quickly blanket large regions of the Earth's surface. The fact that it can now be quantitatively interpreted and understood paves the way for these kinds of systems being incorporated into existing seismological frameworks for earthquake detection and location, tomography, and natural hazard monitoring.\\


\section{Acknowledgements}
We are grateful to the staff and management at OTE, including Christina Lessi, Dimitris Polydorou, Diomidis Skalistis, Petros Vouddas, and Ioannis Chochliouros, for allowing the use of their pre-existing infrastructure, for assisting with the fiber links, and for housing both systems for the duration of the experiments. We are also grateful to Athena Chalari and others from Silixa for on-site support with the DAS interrogator. Additional field support was provided by Jonas Igel and Sara Klaasen. Maps were made with PyGMT \cite{pygmt2021}, and satellite imagery is from Google Earth.\\

\section{Open Research}
Data from both systems is available, along with Python Jupyter notebooks to reproduce the figures. (Permanent data archiving is underway via the ETH University Research Collection repository. In the meantime, data and scripts are available at the polybox link (similar to dropbox): \url{https://polybox.ethz.ch/index.php/s/dNiaWyyMz22zGiM}.)

\bibliography{references.bib}

\end{document}